\documentclass[%
 aip,
 amsmath,amssymb,
 reprint,%
]{revtex4-1}

\usepackage{graphicx}
\usepackage{dcolumn}
\usepackage{bm}

\usepackage[utf8]{inputenc}
\usepackage[T1]{fontenc}
\usepackage{mathptmx}
\usepackage{etoolbox}

\makeatletter
\def\@email#1#2{%
 \endgroup
 \patchcmd{\titleblock@produce}
  {\frontmatter@RRAPformat}
  {\frontmatter@RRAPformat{\produce@RRAP{*#1\href{mailto:#2}{#2}}}\frontmatter@RRAPformat}
  {}{}
}%
\makeatother
\begin{document}

\preprint{AIP/123-QED}

\title[Bird's-Eye View of Primitive Chaos]{Bird's-Eye View of Primitive Chaos}
\author{Yoshihito Ogasawara}
 \affiliation{Advanced Institute for Complex Systems, Comprehensive Research Organization, Waseda University.}
 \altaffiliation[Also at ]{Miyoshi Gokin Kogyo Co. Ltd.}
\email{yoshihito.ogasawara@yamatogokin.com}


\date{\today}

\begin{abstract}
The notion of primitive chaos was proposed [J. Phys. Soc. Jpn. {\bf 79}, 15002 (2010)] as a notion closely related to the fundamental problems of science itself such as determinism, causality, free will, predictability, and irreversibility. In this article, we introduce the notion of bird's-eye view into the primitive chaos, and we find a new hierarchic structure of the primitive chaos. This means that if we find a chaos in a real phenomenon or a computer simulation, behind it, we can clearly realize the possibility of tremendous varieties of chaos in the hierarchic structure unless we can see them visually.
\end{abstract}

\maketitle

\begin{quotation}
The notion of primitive chaos was proposed as a notion closely related to the fundamental problems of science itself such as determinism, causality, free will, predictability, and irreversibility. Exploring the existence of the primitive chaos, we attained two contrast notions, nondegenerate Peano Continuum and Cantor set, together with the notions of part and whole, hierarchy, coarse graining, self-similarity, analogy, and logic. Then, it was revealed that this primitive chaos is literally a primitive chaos in such a sense that this primitive chaos becomes a conventional chaos under additional conditions. In this study, we introduce a new notion, bird's-eye view, into the primitive chaos.
\end{quotation}

\section{\label{sec:level1}Introduction}

Since the notion of primitive chaos was proposed,\cite{O2010,O2011} it has been studied as a notion closely related to the fundamental problems of science itself.\cite{O2010,O2011,O2012,O2014,O2015} \\

{\noindent\bf Definition 1.}~{\it If a set $X$, the family of nonempty subsets of $X$, $\{X_\lambda,~\lambda\in\Lambda\}$, and the family of maps, $\{f_{X_\lambda}:X_\lambda\to X,~\lambda\in\Lambda\}$, satisfy the following property (P), $(X,~\{X_\lambda,~\lambda\in\Lambda\},~\{f_{X_\lambda},~\lambda\in\Lambda\})$ is called a primitive chaos.
\begin{itemize}
\item[(P)] For any infinite sequence $\omega_0,~\omega_1,~\omega_2,\ldots$, there exists an initial point $x_0\in\omega_0$ such that
\begin{eqnarray}
f_{\omega_0}(x_0)\in\omega_1,~f_{\omega_1}(f_{\omega_0}(x_0))\in\omega_2,\ldots,\label{P}
\end{eqnarray} where each $\omega_i$ is an element of the family $\{X_\lambda,~\lambda\in\Lambda\}$.
\end{itemize}}
\vspace{4mm}

\noindent In the primitive chaos, each set $X_\lambda$ implies an event or a selection, each sequence $(\omega_n)_{n=0}^\infty$ implies a time series, and each map $f_{X_\lambda}$ implies a law or causality.\cite{O2010,O2011,O2012,O2014,O2015}

Accordingly, if a phenomenon described by a time series $(\omega_n)_{n=0}^\infty$ is obtained even stochastically, it can be traced deterministically, in the primitive chaos. This fact reminds us of the fundamental problem: ``What is the physical laws themselves?" or ``What is the determinism?"

Regarding each set $X_\lambda$ as a selection, we can see that the time series $(\omega_n)_{n=0}^\infty$ is a history of choosing selections. Then, even if the history is chosen of our own free will, it can be traced deterministically (like a destiny). This fact reminds us of the problem: ``What is free will?"

Regarding $\omega_n$ as a present event, we can see that any past,
$$
\omega_0,~\omega_1,~\omega_2,~\ldots,~\omega_{n-1},
$$
 can be traced or recognized by the causality $\{f_{X_\lambda},~\lambda\in\Lambda\}$. However, even the next event $\omega_{n+1}$ (future) is unpredictable; in fact, for any $\omega_{n+1}$, the time series,
$$
\omega_0,~\omega_1,~\omega_2,~\ldots,~\omega_n,~\omega_{n+1},
$$
is described by the causality, and it means that we can unpredict which $\omega_{n+1}$ is.
 This fact reminds us of the problem of time asymmetry or irreversibility. 

Here, it is noted that the notion of the primitive chaos is described simply by sets and maps, without any structures such as topological structures, algebraic structures, and the structures of measure theory.\cite{Of} Namely, the primitive chaos is a notion raising the fundamental problems in science itself in a primitive or essential form; this is quite different from discussions about concrete maps such as the tent map $\varphi:[0,1]\to[0,1],~x\mapsto\min\{2x,2(1-x)\}$, although the primitive chaos is extracted from the property of the concrete maps.\cite{O2010} Then, owing to this extraction, the notion of the primitive chaos has the ability of exploring the fundamental problems from a more profound viewpoint.

On the other hand, under some conditions, the primitive chaos leads to 
the characteristic properties of the conventional chaos,\cite{O2012,RLD} such as the existence of a nonperiodic orbit, the existence of the periodic point whose prime period is $n$ for any $n\in\mathbb{N}$, the existence of a dense orbit, the density of periodic points, sensitive dependence on initial conditions, and topological transitivity. In this sense, this primitive chaos is literally a primitive chaos, where the property (P) plays an essential role in the demonstration of the above properties of the conventional chaos. Further, the additional conditions are natural in such a sense that the concrete maps such as the tent map satisfy these conditions naturally.

Then, we explored the essential qualities of guaranteeing the existence of the primitive chaos from a topological viewpoint. Here, it is noted that topology has the ability to determine our method of viewing;\cite{Lewin,Thom,YO,Epis} for example, we can see that the closed interval $[0,1]$ is totally disconnected under the discrete topology, although it is of course a connected space under the standard topology induced by the Euclid metric. That is, the discussion from a topological viewpoint implies the discussion on our method of viewing. In addition, the topological discussion is general or universal. In fact, a topological space can represent a matter itself in the  Euclid space $\mathbb{R}^3$ and subspaces in the spatiotemporal space, in the phase space, in the configuration space, in the Hilbert space, and in the function space. That is, topology can describe universal properties regardless of their individuality.

Then, we attained two characteristic topological notions, nondegenerate Peano continuum and Cantor set, together with the notions of part and whole,\cite{O2011,O2014} hierarchy,\cite{O2011,O2014,O2015} coarse graining,\cite{O2011,O2015} self-similarity,\cite{O2011,O2014} analogy,\cite{O2011} and logic.\cite{O2014,Stone} A nondegenerate set means a set consisting of more than one point. A Peano continuum is a locally connected continuum, and a continuum is a nonempty compact connected metric space. The nondegenerate Peano continuum is a topologically extracted and essential notion such that it has many examples\cite{SBN,hyper} such as arcs, $n$-cells, $n$-spheres, toruses, solid toruses, trees, graphs, nondegenerate dendrites,\cite{O2010,O2011} and Hilbert cubes. 

A Cantor set is a space homeomorphic to the Cantor middle-third set, and it is known that a space is a Cantor set if and only if it is a zero-dimensional perfect compact metrizable space.\cite{Cant} A topological space is zero-dimensional provided that there is a base for its topology such that each element of the base is clopen (closed and open), and a topological space is perfect provided that it contains no isolated points. The Cantor set is also a topologically extracted and essential notion, quite differently from the special set, the Cantor middle-third set.\cite{O2014}

Then, if $X$ is the nondegenerate Peano continuum or the Cantor set, $X$ guarantees not only the existence of the primitive chaos, but also the existence of its infinite variety;\cite{O2010,O2014} $X$ guarantees the infinite varieties of families $\{X_\lambda\}$\cite{Boolean} and $\{f_{X_\lambda}\}$\cite{map} such that $(X,\{X_\lambda\},\{f_{X_\lambda}\})$ is a primitive chaos, and each $X_\lambda$ guarantees the infinite varieties of families $\{X_{\lambda_\mu}\}$ and $\{f_{X_{\lambda_\mu}}\}$ such that $(X_\lambda,\{X_{\lambda_\mu}\},\{f_{X_{\lambda_\mu}}\})$ is a primitive chaos. This procedure can be repeated, and thus the hierarchical structure of the primitive chaos is constructed with infinite variety. Then, through the process, there emerge the notions of part and whole, hierarchy, coarse graining, self-similarity, analogy, and logic.

On the other hand, it is known that we are surrounded by diverse chaotic behaviors. The above discussion can provide an answer of why we are.\cite{O2012,O2014} In fact, the nondegenerate Peano continuum and the Cantor set is essential notions which have a great many examples, and they guarantee the existence of infinite variety of the primitive chaos leading to the conventional chaotic behaviors under the natural conditions. 

Furthermore, we can see the contrast between these notions, because the nondegenerate Peano continuum is characterized by its continuum and the Cantor set is characterized by its zero-dimensionality. This contrast reminds us of the contrasting aspects of matter from a macroscopic viewpoint and a microscopic viewpoint. Then, recalling that the notion of the primitive chaos is closely related to the problem ``What is the physical laws themselves?", the notions of continuity (continuum) and discreteness (zero-dimensionality) seem to be our intrinsic notions for the method of recognizing phenomena itself.\cite{O2014} In addition, the nondegenerate Peano continuum and the Cantor set are closely related to each other. We can find a Cantor set in a nondegenerate Peano continuum together with the notion of (weak) self-similarity.\cite{CSF1} For any Cantor set $X$ and any nondegenerate Peano continuum $Y$, there exists a decomposition space $\mathcal{D}$ of $X$ such that $\mathcal{D}$ is homeomorphic to $Y$;\cite{Deco} that is, we can regard the nondegenerate Peano continuum as a coarse graining of  the Cantor set.\cite{O2011} This seems to provide such a picture that we can see a microscopic structure in a macroscopic structure as an analogy of macroscopic phenomena, and we can see a macroscopic structure as a coarse graining of a microscopic structure.  

Now, we can recognize that the notion of the primitive chaos, which has fundamental meanings in science itself, leads to fertile notions and structures. This seems to have the possibility of providing a new method of viewing our world.\cite{Hanson,Kuhn,Fey,Mach,Epis,mono,sekai} In this study, let us add a new view, a bird's-eye view, to it.

\section{Bird's-Eye View of Primitive Chaos}

At first, let us recall a topological notion, Vietoris topology.\cite{hyper}\\

{\noindent\bf Definition 2.}~{\it Let $(X,\tau)$ be a topological space, and $CL(X)$ be the set of all nonempty closed subsets of $X$. Then, the Vietoris topology $\tau_V$ for
$CL(X)$ is the smallest topology which has the following conditions 
$$
\{A\in CL(X);~A\subset G\}\in\tau_V
$$
whenever $G\in\tau$, and
$$
\{A\in CL(X);~A\subset F\}
$$
is $\tau_V$-closed whenever $F$ is $\tau$-closed.}\\

Here, 
$$
\mathcal{B}_V=\{<G_1,\ldots,G_n>;~G_i\in\tau~for~each~i~and~n~(<\infty)\}
$$
is a base for $\tau_V$, where
\begin{eqnarray*}
<S_1,\ldots,S_n>
&=&\{A\in CL(X);~A\subset S_1\cup\cdots\cup S_n\cr
&&~and~A\cap S_i\ne\emptyset~for~each~i\}
\end{eqnarray*}
for any finitely many subsets $S_1,\ldots,S_n$ of $X$.

The meaning of the Vietoris topology is revealed through the notion of the following Hausdroff metric.\cite{hyper}\\

{\noindent\bf Definition 3.}~{\it Let $(X,d)$ be a metric space,\cite{bounded} and $C(X)$ be the set of all nonempty compact subsets of $X$. Then, the Hausdorff metric $H_d$ for $C(X)$ is defined by
\begin{eqnarray*}
H_d(A,B)=\inf\{r>0;~A\subset N_d(r,B),~B\subset N_d(r,A)\},\cr
N_d(r,A)=\{x\in X;~d(x,A)<r\},\cr
d(x,A)=\inf\{d(x,a);~a\in A\}.
\end{eqnarray*}}
 
This Hausdorff metric $H_d$ literally satisfies the definition of metric, while the distance of $A$ and $B$, $\inf_{a\in A,~b\in B}d(a,b)$, does not satisfy the definition of metric. Namely, $(C(X),H_d)$ is a new metric space constructed from the original metric space $(X,d)$. 

Here, since $X$ is Hausdorff, $C(X)$ is the subset of $CL(X)$ (if $X$ is compact, $C(X)=CL(X)$). Then, the subspace $(C(X),\tau_V|C(X))$ of $(CL(X),\tau_V)$ is identical with $(C(X),\tau_{H_d})$ indued by the metric $H_d$.\cite{Vietoris} It is noted that the definition of the Vietoris topology unreqires the metrizability of $X$; the Vietoris topology can be defined for any topological space $X$. In this sense, the notion of the Vietoris topology $\tau_V$ is more extracted than the topology $\tau_{H_d}$.\cite{recall}

Then, let us consider the meaning of the space $(C(X),\tau_{H_d})$. Let a metric space $(X,d)$ be the the Euclid space $(\mathbb{R}^3,d_3)$, where $d_3$ is the Euclid metric. Let each compact set $A\in C(\mathbb{R}^3)$ be regarded as a matter in $\mathbb{R}^3$. Let us consider such a situation that an undeformed matter (i.e., a rigid body) moves in $\mathbb{R}^3$. Then,  its moving distance in $\mathbb{R}^3$ is identical with its Hausdorff metric in $C(\mathbb{R}^3)$. In addition, let us consider that a matter described by a closed ball with a center $a$,
\begin{eqnarray}
B_{d_3}(a,r)=\{x\in\mathbb{R}^3;~d_3(x,a)\le r\},
\end{eqnarray}
changes its radii, that is, it deforms from
\begin{eqnarray}
A_1=B_{d_3}(a,r_1)
\end{eqnarray}
to
\begin{eqnarray}
A_2=B_{d_3}(a,r_2).
\end{eqnarray}
Then, $H_{d_3}(A_1,A_2)$ is identical with the difference $|r_1-r_2|$ of these radius. Namely, if a matter swells (or shrinks) from $A_1$ to $A_2$ in $\mathbb{R}^3$, its difference is identical with $H_{d_3}(A_1,A_2)$. In addition, particular if $A_1$ and $A_2$ are singletons $\{a_1\}$ and $\{a_2\}$, respectively, $H_{d_3}(A_1,A_2)$ is identical with $d_3(a_1,a_2)$. Accordingly, in this sense, the Hausdorff metric $H_{d_3}$ is a metric naturally extended from the original Eucrid metric $d_3$ (as to $\mathbb{R}^1$ and $\mathbb{R}^2$, the same discussions hold). 

However, it must be noted that $X$ and $C(X)$ are the discussions in completely different levels. In fact, $A\in C(X)$ is a point in the set $C(X)$, while $A$ is a set in $X$. A matter $A\in C(\mathbb{R}^3)$ is of course a subspace of the Euclid space $(\mathbb{R}^3,d_3)$, but it is a point in the space $(C(\mathbb{R}^3),\tau_{H_{d_3}})$. This corresponds to such a situation that a matter is seen at distance as if a bird in the sky sees down our world. Therefore, in this study, let us regard the space $(C(X),\tau_{H_d})$ as the bird's-eye view of the original space $(X,\tau_d)$, while it is called a hyperspace mathematically.\cite{hyper}

With the aid of the notion of the bird's-eye view, for instance, we can consider the following phenomenon. Let us suppose that a matter moves and deforms in the space $\mathbb{R}^3$, and its situation is described by a time series,
\begin{eqnarray}
A_0,~A_1,~A_2,~\ldots~,\label{A}
\end{eqnarray}
where each $A_i$ is an element of $C(\mathbb{R}^3)$. In this time series, if we can find a law or causality,
\begin{eqnarray}
T:C(\mathbb{R}^3)\to C(\mathbb{R}^3),\label{T}
\end{eqnarray}
the time series is represented by
\begin{eqnarray}
A_0,~T(A_0),~T(T(A_0)),~\ldots~.\label{TA}
\end{eqnarray}
Then, we can discuss this phenomenon in the metric space $(C(X),H_d)$ as if a bird in the sky sees down our world, by regarding the matters in $\mathbb{R}^3$ as points in $C(X)$. That is, we can describe the phenomenon itself in the higher level, without simplification such as the center of gravity.

Now, let us recall the discussion of the primitive chaos. Exploring the nature of the primitive chaos, we attained two characteristic notions, nondegenerate Peano continuum and Cantor set, together with fertile structures. This drives us to a question ``What is the relation of the bird's-eye view and these spaces", and we are led to the following theorems.\\

{\noindent\bf Theorem 1.\cite{Peano}}~{\it Let $(X,d)$ be a nondegenerate Peano continuum, and then $(C(X),H_d)$ is also  a nondegenerate Peano continuum.}\\

{\noindent\bf Theorem 2.\cite{Cantor}}~{\it Let $(X,d)$ be a Cantor set, and then $(C(X),H_d)$ is also a Cantor set.}\\

Namely, if $X$ is the nondegenerate Peano continuum or the Cantor set which generates infinite variety of the primitive chaos, $C(X)$ is also the space which generates it. For instance, let us suppose that $X$ is a nondegenerate Peano continuum such as arcs, $n$-cells, $n$-spheres, toruses, solid toruses, trees, graphs, nondegenerate dendrites, and Hilbert cubes. Then, there exists the infinite varieties of families $\{X_\lambda\}$ and $\{f_{X_\lambda}\}$ such that $(X,\{X_\lambda\},\{f_{X_\lambda}\})$ is a primitive chaos, where each $X_\lambda$ is a nondegenerate Peano subcontinuum of $X$ and thus, $\{X_\lambda\}$ is a subset of $C(X)$. Then, since the bird's-eye view $(C(X), H_d)$ is also a nondegenerate Peano continuum, there exists the infinite varieties of families $\{C(X)_\mu\}$ and $\{f_{C(X)_\mu}\}$ such that $(X,\{C(X)_\mu\},\{f_{C(X)_\mu}\})$ is a primitive chaos. That is, for any infinite sequence $\Omega_0,~\Omega_1,~\Omega_2,\ldots$, there exists an initial point $A_0\in\Omega_0$ such that
\begin{eqnarray}
f_{\Omega_0}(A_0)\in\Omega_1,~f_{\Omega_1}(f_{\Omega_0}(x_0))\in\Omega_2,\ldots,
\end{eqnarray} 
where each $\Omega_i$ is an element of the family $\{C(X)_\mu\}$. Then, under additional conditions, the primitive chaos $(X,\{C(X)_\mu\},\{f_{C(X)_\mu}\})$ becomes a conventional chaos in the level, the bird's-eye view $C(X)$. 

Furthermore, since $X$ is compact and $C(X)$ is a Hausdorff space in addition to the existence of a continuous map from $X$ onto $C(X)$,\cite{O2010} there exists a decomposition space $\mathcal{D}(X)$ of $X$ such that $\mathcal{D}(X)$ is homeomorphic to $C(X)$; that is,  we can regard the bird's-eye view $C(X)$ as the coarse graining of the original space $X$.\cite{O2011,YO}

In addition, we can consider the bird's-eye view of $C(X)$, that is,  the bird's-eye view $C(C(X))$ of  the bird's-eye view $C(X)$ of $X$, such that it has the existence of infinite variety of the primitive chaos, $(C(C(X)),~\{C(C(X))_\gamma\},~\{f_{C(C(X))_\gamma}\})$ (In addition, since each $X_\lambda$ and $C(X)_\mu$ are also nondegenerate Peano continua, we can consider the bird's eye view $C(X_\lambda)$ and $C(C(X)_\mu)$, respectively). Then, since this procedure can be repeated, we can see a new hierarchic structure, each element of which guarantees the existence of infinite variety of the primitive chaos. Then, all of the primitive chaos can become the conventional chaos under natural conditions. Also for a Cantor set, the same discussions hold.

\section{Conclusions}

We introduce the notion of the bird's-eye view, and we find a new hierarchic structure of the primitive chaos. Also in the bird's-eye view, under additional conditions, the primitive chaos becomes a conventional chaos. Then, it is known that we are surrounded by diverse chaotic behaviors. If the chaos which we find is guaranteed by the Peano continuum or the Cantor set, behind the chaos, there exist the infinite varieties of the primitive chaos leading the conventional chaos under natural conditions in the bird's-eye view, the bird's-eye view of the bird's-eye view, the bird's-eye view of the bird's-eye view of the bird's-eye view and so on. In the high level of the hierarchic structure of the bird's-eye view, even if the primitive chaos becomes the conventional chaos, we can no longer see the chaotic behavior visually as a real phenomenon or a computer simulation. However, with the aid of the topology which has the ability of discussing the method of viewing, we can now realize the possibility of  the existence of infinite variety of the (primitive) chaos in the hierarchic structure. 

Namely, if we find a chaos in a real phenomenon or a computer simulation, behind it, we can realize the possibility of tremendous varieties of chaos in the hierarchic structure unless we can see them visually. This fact provides a totally new method of viewing our world.



\begin{thebibliography}{9}
\bibitem{O2010} Y. Ogasawara, J. Phys. Soc. Jpn. {\bf 79}, 15002 (2010).
\bibitem{O2011} Y. Ogasawara and S. Oishi, J. Phys. Soc. Jpn. {\bf 80}, 67002 (2011).
\bibitem{O2012} Y. Ogasawara and S. Oishi, J. Phys. Soc. Jpn. {\bf 81}, 103001 (2012).
\bibitem{O2014} Y. Ogasawara and S. Oishi, J. Phys. Soc. Jpn. {\bf 83}, 14001 (2014).
\bibitem{O2015} Y. Ogasawara, J. Phys. Soc. Jpn. {\bf 84}, 64007 (2015).



\bibitem{Of} Of course, these structures are also described only by sets and maps. Actually, we may say that the primitive chaos itself is a new structure. 

\bibitem{RLD} R. L. Devaney, {\it An Introduction to Chaotic Dynamical Systems} (Westview Press, Colorado, 2003).\label{RLD}

\bibitem{Thom} R. Thom, {\it Stabilit\'e Structurelle et Morphog\'en\`ese} (InterEditions, Paris,
1977) [in French].
\bibitem{Lewin} K. Lewin: {\it Principles of Topological Psychology}, transl. F. Heider and G. M. Heider (McGraw-Hill, New York, 1936).

\bibitem{YO} Y. Ogasawara, {\it Mono no Mikata Toshiteno Isokukanron Nyumon} (Baifukan, Tokyo, 2011) [in Japanese].

\bibitem{Epis} Y. Ogasawara, Epistemology and Mind Sciences {\bf 3}, 6 (2021) [in Japanese]; Y. Ogasawara, Epistemology and Mind Sciences {\bf 3}, 48 (2021) [in Japanese].


\bibitem{Stone}  Any Cantor set is a Stone space well-known by the Stone's representation theorem for the
Boolean algebra.

\bibitem{SBN} S. B. Nadler, Jr., {\it Continuum Theory} (Marcel Dekker Inc., New York, 1992).
\bibitem{hyper} A. Illanes and S. B. Nadler, Jr., {\it Hyperspaces} (Marcel Dekker Inc, New York, 1999).

\bibitem{Cant} A. Illanes and S. B. Nadler, Jr., {\it Hyperspaces} (Marcel Dekker Inc, New York, 1999), p. 65.

\bibitem{Boolean} If $X$ is a Cantor set, each $X_\lambda$ implying an event or a selection is a clopen set in $X$; the set of all clopen sets of a topological space is a typical example of the Boolean algebra.\cite{O2014} 

\bibitem{map} If $X$ is a nondegenerate Peano continuum, for each map $f_{X_\lambda}$,  for any positive integer $m$, we can assign $m$ points in advance.\cite{O2010}  This reminds us of the fact that we tend to regard a law as universal, although any law can be verified ony finitely, in principle.\cite{O2014, Hume} 

\bibitem{CSF1}A. Kitada and Y. Ogasawara, Chaos Solitons Fractals 24, 785 (2005); A. Kitada and Y. Ogasawara, Chaos Solitons Fractals 25, 1273 (2005).

\bibitem{Deco} Note that the Cantor set $X$ is compact, the nondegenerate Peano continuum $Y$ is Hausdorff, and there exists a continuous map from $X$ onto $Y$.\cite{O2014}

\bibitem{Hanson}N. R. Hanson, Patterns of Discovery: An Inquiry into the Conceptual Foundations of Science (Cambridge University Press, Cambridge, U.K., 1958).
\bibitem{Kuhn} T. S. Kuhn, The Structure of Scientific Revolutions (University of Chicago Press, Chicago, IL, 1970).
\bibitem{Fey}P. Feyerabend, Against Method: Outline of an Anarchistic Theory of Knowledge (Humanities Press, London, 1975).
\bibitem{Mach} E. Mach, {\it Die Analyse der Empfindungen und das Verh\"altnis des Physischen zum Psychischen} (Gustav Fischer, Jena, 1922) [in German].
\bibitem{mono} S. Ohmori, {\it Mono to Kokoro} (University of Tokyo Press, Tokyo, 1980) [in Japanese].

\bibitem{sekai}  T. Nagel, {\it The View from Nowhere} (Oxford University Press, New York, 1986).

\bibitem{bounded} If $X$ is bounded, $H_d$ can be defined for $CL(X)$. 

\bibitem{Vietoris} A. Illanes and S. B. Nadler, Jr., {\it Hyperspaces} (Marcel Dekker Inc, New York, 1999), p. 16.

\bibitem{recall}  The definition of the primitive chaos also unrequires metrizability. 

\bibitem{Peano} A. Illanes and S. B. Nadler, Jr., {\it Hyperspaces} (Marcel Dekker Inc, New York, 1999), p. 89.

\bibitem{Cantor} S. B. Nadler, Jr., {\it Continuum Theory} (Marcel Dekker Inc, New York, 1992), p. 114.

\bibitem{Hume} D. Hume, {\it A Treatise of Human Nature} (J.M. Dent \& Sons Ltd., London; E.P. Dutton \& Co., Inc., New York, 1911).


\end{thebibliography}
\end{document}